# Towards a Theory of Evolution as Multilevel Learning

Vitaly Vanchurin[1,2,*], Yuri I. Wolf[1], Mikhail I. Katsnelson[3], Eugene V. Koonin[1,*]

[1]National Center for Biotechnology Information, National Library of Medicine, Bethesda, MD 20894;

[2]Duluth Institute for Advanced Study, Duluth, Minnesota, 55804, USA;

[3]Radboud University, Institute for Molecules and Materials, Nijmegen, 6525AJ, The Netherlands

*For correspondence: <u>vitaly.vanchurin@gmail.com</u>, <u>koonin@ncbi.nlm.nih.gov</u>

**Abstract**

We apply the theory of learning to physically renormalizable systems in an attempt to develop a theory of biological evolution, including the origin of life, as multilevel learning. We formulate seven fundamental principles of evolution that appear to be necessary and sufficient to render a universe observable and show that they entail the major features of biological evolution, including replication and natural selection. These principles also follow naturally from the theory of learning. We formulate the theory of evolution using the mathematical framework of neural networks, which provides for detailed analysis of evolutionary phenomena. To demonstrate the potential of the proposed theoretical framework, we derive a generalized version of the Central Dogma of molecular biology by analyzing the flow of information during learning (back-propagation) and predicting (forward-propagation) the environment by evolving organisms. The more complex evolutionary phenomena, such as major transitions in evolution, in particular, the origin of life, have to be analyzed in the thermodynamic limit, which is described in detail in the accompanying paper.

**Significance statement**

Modern evolutionary theory gives a detailed quantitative description of microevolutionary processes that occur within evolving populations of organisms, but evolutionary transitions and emergence of multiple levels of complexity remain poorly understood. Here we establish correspondence between the key features of evolution, renormalizability of physical theories and learning dynamics, to outline a theory of evolution that strives to incorporate all evolutionary processes within a unified mathematical framework of the theory of learning. Under this theory, for example, natural selection readily arises from the learning dynamics, and in sufficiently complex systems, the same learning phenomena occur on multiple levels or on different scales, similar to the case of renormalizable physical theories.

# 1. Introduction

What is life? If this question is asked in the scientific rather than in the philosophical context, a satisfactory answer should assume the form of a theoretical model of the origin and evolution of complex systems that are identified with life, from inanimate matter. NASA has operationally defined life as follows: "Life is a self-sustaining chemical system capable of Darwinian evolution" (1, 2). Apart from the insistence on chemistry, long-term evolution that involves (random) mutation, diversification and adaptation is, indeed, an intrinsic, essential feature of life that is not apparent in any other natural phenomena. The problem with this definition, however, is that natural (Darwinian) selection itself appears to be a complex rather than an elementary phenomenon (3). In all evolving organisms we are aware of, for natural selection to kick off and to sustain long-term evolution, an essential condition is the replication of a complex digital information carrier (a DNA or RNA molecule). The fidelity of replication must be sufficiently high to provide for the differential replication of emerging mutants and survival of the fittest ones (this replication fidelity level is often referred to as Eigen threshold) (4). In modern organisms, accurate replication is ensured by highly complex molecular machineries that include not only replication and repair enzymes, but also the entire metabolic network of the cell that provides energy and building blocks for replication. Thus, the origin of life is a typical chicken-and-egg problem (or a Catch-22): sufficiently accurate replication is essential for evolution but the mechanisms underlying such a replication process are themselves products of complex evolutionary processes (5, 6).

Because the replication capacity of living organisms is itself a product of evolution, a solution to the problem of the origin of life is to be sought outside the traditional framework of evolutionary biology. Modern evolutionary theory, steeped in population genetics, takes a detailed and, arguably, largely satisfactory account of microevolutionary processes, that is, evolution of allele frequencies in a population of organisms under selection and random genetic drift (7, 8). However, the population genetic theory has little to say about the actual historical development of life, especially, about macroevolution that involves emergence of new levels of biological complexity, and nothing at all about the origin of life.

The crucial feature of biological complexity is its hierarchical organization. Indeed, multilevel hierarchies permeate biology. From small molecules to macromolecules; from macromolecules to functional complexes, subcellular compartments, and cells; from unicellular organisms to communities, consortia and multicellularity; from simple multicellular organisms to highly complex forms with differentiated tissues; from organisms to communities and eventually to eusociality and to complex biocenoses involved in biogeochemical processes on the planetary scale. All these distinct levels jointly constitute the hierarchical organization of the biosphere. Understanding the origin and evolution of this hierarchical complexity can be considered one of the principal goals of biology.

In large part, evolution of the multilevel organization of biological systems appears to be driven by solving optimization problems, which entails conflicts, or trade-offs between the optimization criteria at different levels or scales, leading to frustrated states, in the language of physics (9-11). Two notable cases in point are the parasite-host arm race that permeates biological evolution and makes major contributions to the diversity and complexity of life forms (12-15), and multicellular organization of complex organisms, where the tendency of individual cells to reproduce at the highest possible rate is countered by the control of cell division imposed at the level of a multicellular organism (16, 17).

Two tightly linked but distinct, fundamental concepts that lie effectively outside the canonical narrative of evolutionary biology address evolution of biological complexity: major transitions in evolution (MTE) (18-20) and multilevel selection (MLS) (21-26). Each MTE involves the emergence of a new level of organization, often described as an evolutionary transition in individuality. A clear-cut example is the evolution of multicellularity, whereby a new level of selection emerges, namely, selection among ensembles of cells rather than among individual cells. Importantly, multicellular life forms (even counting only complex organisms with multiple cell types) evolved on many independent occasions during the evolution of life (27, 28), strongly suggesting that emergence of new levels of complexity is a major evolutionary trend rather than a rare event occurring by chance.

The MLS is often perceived as a controversial concept, presumably, because of the link to the long-debated subject of group selection (26, 29). However, as a defining component of MTE, multilevel selection appears to be indispensable. A proposed general mechanism behind the MTE, formulated by analogy with the physical theory of the origin of patterns, for example, in glass-like systems, involves competing interactions at different levels and the frustrated states such interactions cause (11). In the physical theory of spin glasses, frustrations result in non-ergodicity and enable formation and persistence of long-term memory, that is, history (30, 31). By contrast, ergodic systems have no true history because they reach all possible states during their evolution (at least in the large time limit), and thus, the only content of quasi-history of such systems is the transition from less probable to more probable states for purely combinatorial reasons, that is, entropy increase (32). As emphasized by Schrödinger in his seminal book (33), even if only in general terms because no adequate theory existed at the time, life is based on "negentropic" processes, and frustrations at different levels are necessary for these processes to set off and persist (11). Conflicting interactions and frustrated states in biological systems are intimately linked to solving optimization problems, which involves multiple temporal and spatial scales. Again, the most obvious case in point seems to be the origin of multicellularity, where different selective factors operate at different levels or scales: selection for the rate of proliferation at the level of individual cells vs selection for cell division control at the level of multicellular ensembles. Similarly, at a higher plane of organization, selection affecting individuals clashes with the selection at the level of groups, communities, or societies, for example, in social insects. At a lower level, the frustrations emerge between the selection affecting "selfish" individual genes and genetic elements, such as transposons, and the selection for the entire genome as a structured collection of genes (23). In each MTE, by definition, selection at the higher level supersedes selection at the lower level (19), but the complexity enabled by this hierarchy of selective factors comes at the cost of elaborate and error-prone mechanisms that keep in check the lower-level units of selection.

The origin of the first cells, which can and probably should be equated with the origin of life, was the first and most momentous transition at the onset of biological evolution, and as such, is outside the purview of evolutionary biology *sensu stricto*. Arguably, theoretical investigation of the origin of life can be feasible only within the framework of an envelope theory that would incorporate biological evolution as a special case (34, 35). It is natural to envisage such a theory as encompassing the non-ergodic processes occurring throughout the history of the universe, with the origin and evolution of life being a special case emerging under conditions that remain to be investigated and defined.

Here, in pursuit of a maximally general theory of evolution, we adopt the formalism that was originally developed in the theory of machine learning (36). It has to be emphasized that learning here is perceived in the maximally general sense, as an objective process that might occur in all evolving systems, including but not limited to biological ones (37). As such, the analogy between learning and selection appears obvious: both types of processes involve trial-and-error and acceptance or rejection of the results based on some formal criteria; in other words, both are optimization processes (21, 38, 39). Here we assess how far this analogy extends, by establishing the correspondence between the key features of biological evolution and concepts as well as the mathematical formalism of learning theory. In particular, we make the case that loss function that is central to the learning theory can be usefully and generally employed as the equivalent of the fitness function in the context of evolution. We exploit the mathematical framework of the theory of learning (36) to sketch a theory of evolution. Our original motivation was to explain major features of biological evolution from more general principles of physics. However, after formulating such principles and embedding them within the mathematical framework of learning, we find that the theory can potentially apply to the entire history of the evolving universe (37) including physical processes that have been taking place since the Big Bang and chemical processes that directly antedated and set the stage for the origin of life. In particular, we show that learning in a complex environment leads to separation of scales, with trainable variables splitting into at least two classes, faster and slower changing ones. Such separation of scales underlies all processes that involve the formation of complex structure in the universe, from the scale of an atom to that of clusters of galaxies. Scale separation occurs during (pre)biological evolution, and we argue that for the emergence of life, at least three temporal scales, which respectively correspond to environmental, phenotypic, and genotypic variables, are essential. In evolving deep learning systems, the slowest-changing variables are digitized and acquire the ability to replicate, resulting in differential reproduction depending on the loss (fitness) function value, which is necessary and sufficient for the onset of evolution by natural selection. The subsequent evolution of life involves emergence of many additional scales, which correspond to MTE. The key biological features of life, namely, MLS, persistence of genetic parasites and programmed death as well as the key physical features of the universe, namely, hierarchy of scale, frequency gaps and renormalizability (40, 41) are among the central propositions of the theory of evolution presented here.

Hereafter we use the term "evolution" to describe the process of temporal changes of living, life-like and prebiotic systems (a.k.a organisms). The more general term "dynamics" refers to temporal processes in other, in particular, physical systems.

At least since the publication of Schrödinger's book, the possibility has been discussed that, although life forms certainly obey the laws of physics, a different class of physical laws uniquely associated with life could exist. Often, this putative physics of life is associated with emergence (42-44), but the nature of the involved emergent phenomena, to our knowledge, has not been clarified until very recently (37) Here we outline a general approach to modeling and studying evolution, in the form of a multilevel learning process, supporting the claim that a distinct type of physical theory, namely, a theory of learning (36, 37), is necessary to investigate the evolution of complex objects in the universe, of which evolution of life is a specific, even if highly remarkable form. A corollary of this approach seems to be that the emergence of the level of complexity characteristic of life is a general trend in the evolution of learning systems.

2. **Fundamental principles of evolution**

In this section, we attempt to formulate the minimal, universal principles that define an observable universe, in which evolution is possible and, perhaps, inevitable. Our analysis began with the consideration of the major features of biological evolution discussed in the next section and proceeded towards the general principles. In this presentation, however, we start from the latter, for the sake of transparency and generality.

So, what are the requirements for a universe to be observable? The possibility to make meaningful observations implies a degree of order and complexity in the observed universe provided by some evolutionary processes, and such evolvability itself seems to be predicated on several fundamental principles. Before formulating these propositions explicitly, we have to emphasize that "observation" as well as "learning" here by no means imply "mind" or "consciousness", but a far more basic requirement. To learn and survive in an environment, a system (or observer) must predict, with some minimal but sufficient degree of accuracy, the response of that environment to various actions and to be able to choose such actions that are compatible with the observer's further existence in that environment. In this sense, any life form is an observer, and even inanimate entities endowed with the ability of feedback reaction qualify as observers. In this, most general sense, observation is a pre-requisite of evolution. We first formulate the basic principles underlying observability and evolvability, and then, give the pertinent comments and explanations.

> P1. **Loss function.** In any evolving system, there exists a loss function of time-dependent variables that is minimized during evolution.
>
> P2. **Hierarchy of scales**. Evolving systems encompass multiple dynamical variables that change on different temporal scales (with different characteristic frequencies).
>
> P3. **Frequency gaps**. Dynamical variables are split among distinct levels of organization separated by sufficiently wide frequency gaps.
>
> P4. **Renormalizability.** Across the entire range of organization of evolving systems, a statistical description of faster-changing (higher frequency) variables is feasible through the slower-changing (lower frequency) variables.
>
> P5. **Extension**. Evolving systems have the capacity to recruit additional variables that can be utilized to sustain the system and the ability to exclude variables that could destabilize the system.
>
> P6. **Replication.** In evolving systems, replication and elimination of the corresponding information processing units can take place on every level of organization.
>
> P7. **Information flow.** In evolving systems, slower-changing levels absorb information from faster-changing levels during learning and pass information down to the faster levels for prediction of the state of the environment and the system itself.

The first principle P1 is of special importance as the starting point for a formal description of evolution as a learning phenomenon. Indeed, the very existence of a loss function implies that the dynamical system of the universe, or simpler, the universe itself is a learning (evolving) system (36). Effectively, here we assume that stability or survival of any subsystem of the universe is equivalent to solving an optimization or learning problem in the mathematical sense and that there is always something to learn. Furthermore, the description of evolution as an optimization or learning problem immediately defines the type of mathematical apparatus that is best suited for its analysis (36). Arnold formulated the main message of Newtonian mechanics in one simple proposition: "It is useful to solve (ordinary) differential equations in physics" (45). Similarly, our first principle, in effect, simply states: It is useful to formulate and solve learning (optimization) problems in the theory of evolution. In a form so general, this principle might appear almost trivial, but as discussed below, it has numerous major implications and corollaries. Arguably, the single most important of such corollaries is that, for solving complex optimization problems dependent on many variables, the best and in fact the only efficient method is selection implemented in various stochastic algorithms (Markov Chain Monte Carlo, stochastic gradient descent, genetic algorithms and more). All evolution can be perceived as an implementation of a stochastic learning algorithm as well.

The remaining principles P2 to P7 provide sufficient conditions for the observers of our type (that is, complex life forms) to evolve within a learning system. In particular, P2, P3 and P4 comprise the necessary conditions for observability of a universe by any observer, whereas P5, P6 and P7 represent the defining conditions for the origin of life of our type (hereafter we omit the qualification for brevity). More precisely, P2 and P3 provide for the possibility of at least a simple form of learning of the environment (fast-changing variables) by an observer (slow-changing variables), and hence the emergence of complex organization of the slow-changing variables. P4 corresponds to the physical concept of renormalizability, or renormalization group (40, 41), whereby the same macroscopic equations, albeit with different parameters, govern processes at different levels or scales, thus limiting the number of relevant variables, constraining the complexity, and allowing for a coarse-grained description. This principle ensures a renormalizable universe capable of evolution and amenable to observation. Together, P2 to P4 define a universe, in which partial or approximate knowledge of the environment, in other words, coarse graining, is both attainable and useful for the survival of evolving systems (observers). Indeed, to use water or food to support our life, there is no need to take into account that it consists of molecules, molecules of atoms, atoms of electrons and nuclei, and so forth. To make the decision to drink or not to drink, it is sufficient to assess a few macroscopic characteristics of the liquid in question, such as color and smell, which can be learned by trial and error. To survive in the macroscopic world, there is no need to study its structure and properties down to the Planck scale (43, 44). In a universe where P4 does not apply, that is, one with non-renormalizable physical laws, what happens at the macroscopic level, will essentially depend on the details of the processes at the microlevel. In a universe where P2 and P3 do not apply, the separation of the macro and micro levels itself would not be apparent. In such a universe, it would be literally impossible to survive without first discovering fundamental physical laws, whereas living organisms on our planet have evolved for billions of years before starting to study quantum physics.

Principles P5, P6 and P7 endow evolving systems with the access to more advanced algorithms for learning and predicting the environment, paving the way for the evolution of complex systems including, eventually, life. These principles jointly underlie the emergence of the crucial phenomenon of selection (46, 47). In its simplest form, selection is for stability and persistence of evolving, learning systems (34, 48). Learning and survival are tightly linked because survival is predicated on the ability of the system to learn the environment, whereas the latter ability depends on the stability of the system on time scales required for learning. Roughly, a system cannot survive in a world where the properties of the environment change faster than the evolving system can learn them. According to P5, evolving systems consume resources (such as food), which themselves could be produced by other evolving systems, to be utilized as building blocks and energy that are required for learning. This principle actualizes Schrodinger's vision (often misunderstood) that "organisms feed on negentropy" (33). Under P6, replication of the carriers of slowly changing variables becomes the basis of long-term persistence and memory in the evolving systems. This principle can be viewed as a learning algorithm built on P3, whereby the time scales characteristic of an individual organism and of consecutive generations are separated. This principle excludes from consideration certain imaginary forms of life, for example, Stanislav Lem's famous Solaris (49). Finally, P7 describes how information flows between different levels in the multilevel learning which gives rise to the concept of a generalized Central Dogma of molecular biology, which is discussed in Sec. 7.

### 3. Fundamental evolutionary phenomena

In this section, we link the fundamental principles of evolution P1-P7 formulated above to the basic phenomenological features of life E1-E10, and seek equivalencies in the theory of learning. The list below is organized by first formulating a biological feature, and then, tracing the connections to the fundamental principles (a), and adding more general comments (b).

E1. **Information-processing units**. Existence of discrete information-processing units (IPUs), that is, self vs non-self differentiation and discrimination, at different levels in organization. All biological entities at all levels of organization, such as genes, cells, organisms, populations, and so on, up to the level of the entire biosphere, possess some degree of self-coherence that separates them, first and foremost, from the environment at large and from other similar-level IPUs.

   a. The existence of IPUs is predicated on the fundamental principles P1-P4 discussed above. The wide range of the temporal scales P2 in dynamical system and gaps separating the scales P3 naturally allow for the separation of slower- and faster-changing components. In particular, renormalizability P4 applies to the hierarchy of IPUs. The statistical predictability of the higher frequencies allows the IPUs to decrease the loss function of the lower frequencies despite the much slower reaction times.
   b. Separation of (relatively) slow-changing pre-biological IPUs from the (typically) fast-changing environment kicked off the most primitive form of pre-biological selection, selection for stability and persistence (survivor bias). More stable, slower-changing IPUs win in the competition and accumulate with time, increasing the separation along the temporal axis as the boundary between the IPUs and the environment grows sharper. Additional key phenomena, such as utilization of available environmental resources P5 and stimulus-response mode of information exchange P7, stem from the flow of matter and information across this boundary and the ensuing separation of internal and external physico-chemical processes. Increasing self vs non-self differentiation, combined with replication of the carriers of slow-changing variables P6 sets the stage for competition between evolving entities and for the onset of the ultimate evolutionary phenomenon, natural selection E6. It has to be emphasized once again that selection is the only efficient method for solving complex optimization problems, to which category the problems faced by evolving IPUs certainly belong.

E2. **Frustration.** All complex, dynamical systems face multi-dimensional and multi-scale optimization problems which generically leads to frustration resulting from conflicting objectives at different scales. This is a key, intrinsic feature of all such systems and a major force driving the advent of increasing, multilevel complexity (11). Frustration is an extremely general physical phenomenon that is by no account limited to biology but rather occurs in much simpler physical systems, such as spin and structural glasses, the behavior of which is determined by competing interactions, so that a degree of complexity is attained (30, 31).

   a. The multi-scale organization of the universe P2 provides the physical foundation for the pervasiveness of frustrated states that typically arise whenever there is a conflict (trade-off) between short- and long-range optimization problems. Frustrated interactions yield multi-well potential landscapes, in which no single state is substantially fitter than numerous other local optima. Multi-parameter and multi-scale optimization of the loss function on such a landscape involves non-ergodic (history-dependent) dynamics, which is characteristic of complex dynamical systems.
   b. IPUs face conflicting interactions starting from the most primitive pre-biological state (11). Indeed, the separation of any type of entities from the environment immediately results in the conflict of permeability: a stronger separation enhances the self vs non-self differentiation, and thus increases the stability of the system, but compromises information and matter exchange with the environment, limiting the potential for growth. In biology, virtually all aspects of the organism architecture and operation are subject to such frustrations, or trade-offs: the conflict between the fidelity and rate of information transmission at all levels; between specialization and generalism; between the individual and population-level benefits; and more. The ubiquity of frustrations and the fundamental impossibility of their resolution in a universally optimal manner are perpetual drivers of the evolution and give rise to evolutionary transitions, attaining levels of complexity that otherwise would be out of reach.

There are two distinct types of frustrations that can be described as spatial and temporal. Spatial frustration is similar to the frustration that is commonly analyzed in condensed matter systems, such as spin glasses(30, 31). In this case, the spatially local and non-local interacting terms have opposite signs so that the equilibrium state is determined by the balance between the terms. In the context of neural networks, a neuron (like a single spin) might have a local objective (such as binary classification of incoming signals) but is also a part of a neural network (like a spin network), which has its own global objective (such as predicting its boundary conditions). For a particular neuron, optimization of the local objective can be in conflict with the global objective, which is the main cause of spatial frustration. Temporal frustration emerges because, in the context of multilevel learning, the same neuron also becomes a part of higher-level IPUs that operate at different temporal scales (frequencies). Then, the optimal state of the neuron with respect to a given IPU operating at a given time scale can differ from the optimal state of the same neuron with respect to another IPU operating at a different time scale (37). Similar to the spatial frustrations, temporal frustrations cannot be completely resolved, but an optimal balance between different spatial and temporal scales is achievable and represents a local equilibrium of the learning system.

E3. **Multi-level hierarchy**. The hierarchy of multiple levels of organization is an intrinsic, essential feature of evolving biological systems that involves both the structure of these systems (genes, genomes, cells, organisms, kin groups, populations, species, communities and more) and the substrate the evolutionary forces act upon.

   a. Renormalizability of the Universe P4 implies that there is no inherently preferred level of organization, for which everything above and below would behave as a homogenous ensemble. Even if some levels of organization come into existence before others (for example, organisms before genes or unicellular organisms before multicellular ones), the other levels will necessarily emerge and consolidate subsequently.
   b. The hierarchy of the structural organization of biological systems was apparent to scholars from the earliest days of science. However, multilevel selection was and remains a controversial subject in evolutionary biology (22, 25, 26). Intuitively and as follows from the Price equation (50), multilevel selection should emerge in all evolving systems as long as the agency of selection possesses a sufficient degree of self vs non-self differentiation. For example, if all organisms of a particular kind (species) exist as populations that are perfectly homogeneous within and are isolated from other, potentially competing biological entities), population-level selection is impossible. If, on the contrary, populations are sufficiently distinct genetically, but interact competitively, population-level selection will inevitably ensue. Clearly, the first case is unrealistic. Evolution of biological systems is driven by conflicting interactions E2 that tend to lead to ever-increasing complexity (11). This trend further feeds the capability and propensity of these systems to form new levels of organization and is associated with evolutionary transitions that involve the advent of new units of selection at multiple levels of complexity.

E4. **Near-optimality**. Stochastic optimization, or the use of stochastic methods for solving optimization problems, is the only feasible approach to complex optimization, but it guarantees neither finding the globally optimal solution nor retention of the optimal configuration when and if it is found. Rather, stochastic optimization tends to rapidly find local optima and keeps the system in their vicinity, sustaining the value of the loss function at a near-optimal level.

   a. According to P1, the dynamics of a learning, that is, self-optimizing system, is defined by a loss function (36, 37). When there is a steep gradient in the loss function, a system undergoing stochastic optimization can rapidly descend in the right direction. However, because of frustrations that inevitably arise from the interactions in a complex system, actual local peaks on the landscape are rarely reached, and the global peak is effectively unreachable. Learning systems tend to get stalled near local saddle points where changes along most of the dimensions either lead "up" or are "flat" in terms of the loss function, with only a small minority of the available moves decreasing the loss function (51).

b. The extant biological systems (cells, multicellular organisms as well as higher level entities, such as populations and communities) are the result of about 4 billion years of the evolution of life on Earth, and as such, are highly, albeit not completely, optimized. As a consequence, the typical distribution of the effects of heritable changes in biological evolution comprises numerous deleterious changes, comparatively rare beneficial changes and common neutral changes, those with fitness effects below the noise level (52). The preponderance of neutral and slightly deleterious changes provides for evolution by neutral genetic drift whereby a population moves on the same level or even slightly downward on the fitness landscape, potentially reaching another region of the landscape where beneficial mutations are available (53, 54).

E5. **Diversity of near-optimal solutions.** Solutions on the loss function landscapes that arise in complex optimization problems, span numerous local peaks of comparable heights.

a. The existence of multiple peaks of comparable heights in the loss function landscapes is a fundamental physical property of frustrated systems (E2), whereas the pervasiveness of frustration itself is a consequence of the multi-scale and multi-level organization of the universe (P2). Frustrated dynamical systems behave in a non-ergodic manner which, from the biological perspective, means that, once separated, evolutionary trajectories diverge, rather than converge. Because most of these trajectories traverse parts of the genotype space with comparable fitness values, competition rarely results in complete dominance of one lineage over the others, but rather generates rich diversity.

b. In terms of evolutionary biology, fitness landscapes are rugged, with multiple adaptive peaks of comparable fitness (55, 56), and a salient trend during evolution is the spread of life forms across multiple peaks as opposed to concentrating on one or few. Evolution pushes evolving organisms to explore and occupy all available niches and try all possible strategies. In the context of machine learning, identical neural networks can start from the same initial state but, for example, under the stochastic gradient descent algorithm, would generically evolve towards different local minima. Thus, the diversity of solutions is a generic property of learning systems. On a more technical level, the diversification is due to the entropy production in the dynamics of the neutral trainable variables (see next section).

E6. **Separation of phenotype from genotype.** This quintessential feature of life embodies the emergence of two distinct (albeit inseparable in known organisms) symmetry-breaking phenomena: i) separation between a dedicated digital information storage media (stable, rarely updatable, tending to distributions with discrete values) and the mostly analog processing devices, and ii) asymmetry of the information flow within the IPUs, whereby genotype provides "instructions" for the phenotype, whereas the phenotype largely loses the ability to update the genotype directly. As such, somewhat paradoxically, the separation between the information storage and processing subsystems is a prerequisite feature that probably emerged early on the path from prebiotic entities to the emergence of life rather than a late-evolved property of living organisms.

a. The separation between phenotype and genotype extends the scale separation on the intra-IPU level, as follows from the fundamental principles P1-P4. Intermediate-frequency components of an IPU (phenotype) buffer the slowest components from direct interaction with the environment (the highest frequency variables), further increasing the stability of the slowest components and making them suitable for long-term information storage. As the temporal scales separate further, the interactions between them change their character. Asymmetric information flow P7 stabilizes the system, enabling long-term preservation of information (genotype), while retaining the reactive flexibility of the faster-changing components (phenotype).

  b. The emergence of the separation between phenotype and genotype is a crucial event in pre-biotic evolution. This separation is prominent in all known as well as hypothetical life forms. Even when the phenotype and genotype roles are fulfilled by molecules that are chemically identical, as in the RNA world scenario of primordial evolution (57, 58), their roles as effectors and information storage devices are sharply distinct. In biological terms, the split is between replicators, the digital information carriers (genomes), and reproducers (44, 45), the analog devices (cells, organisms) that host the replicators, supply them with building blocks P5 and themselves reproduce P6 under the replicators' instruction P7. Although the genotype/phenotype separation is a major staple of life, it is not in itself sufficient to qualify an IPU as a life form (computers and record players, in which the separation between information storage and operational parts is prominent and essential, clearly are not life, even though invented by advanced organisms). The asymmetry of information flow between genotype and phenotype P7 is the most general form of the phenomenon known as the Central Dogma of molecular biology, the unidirectional flow of information from nucleic acids to proteins as originally formulated by Crick (59). This asymmetry is also prominent in other information processing systems, in particular computers. Indeed, von Neumann architecture computers have inherently distinct memory and processing units, with the instruction flow from the former to the latter (60, 61). It appears that any advanced information processing system will be endowed with this property.

E7. **Replication.** Genome replication and sharing of long-term memory. Emergence of long-term digital storage devices (genomes consisting of RNA or DNA; E6) provides for long-term information preservation, facilitates adaptive reactions to changes in the environment, and promotes the stability of IPUs to the point where (at least in chemical systems) it is limited by the energy of the chemical bonds rather than the energy of thermal fluctuations. Obviously, however, as long as this information is confined to a single IPU, it will disappear with the eventually inevitable destruction of that IPU. Should this be the case, other IPUs of similar architecture would need to accumulate a comparable amount of information from scratch to reach the same level of stability. Thus, copying and sharing information is essential for long-term (effectively, indefinite) persistence of IPUs.

  a. The fundamental principle P6 postulates the existence of mechanisms for information copying and elimination. If the genomic information can be replicated, even most primitive sharing mechanisms (such as physical splitting of an IPU under forces of surface tension) would result (even if not reliably) in the emergence of distinct IPUs pre-loaded with information that was amassed by their progenitor(s). This process short circuits learning and allows the information to accumulate at time scales far exceeding the characteristic lifetimes of individual IPUs.

  b. Information copying and sharing is beneficial only if its fidelity exceeds a certain threshold, sometimes called Eigen limit in evolutionary biology (4-6). Nevertheless, in primitive pre-biotic systems, the required fidelity level could have been quite low (62). For instance, even a biased chemical composition of a hydrophobic droplet could enhance the stability of the descendant droplets and thus endow them with an advantage in the selection for persistence. However, once relatively sophisticated mechanisms of information copying and sharing are in place, more precisely, when replicators become information storage devices, the overall stability of the system can increase by many orders of magnitude. To wit, and astonishingly, the only biosphere known to us represents an unbroken chain of genetic information transmission that spans about 4 billion of years, commensurate with the stellar evolution scale.

E8. **Natural selection.** Evolution by natural selection (a.k.a. Darwinian evolution) arises from the combination of all the principles and phenomena described above. The necessary and sufficient conditions for Darwinian evolution to operate are: i) existence of IPUs that are distinct from the environment and from each other (E1); ii) dependence of the stability of an IPU on the information it contains, that is, the phenotype-genotype feedback (E6); iii) ability of IPUs to make copies of embedded information and share it with other IPUs (E7). When these three conditions are met, the relative frequencies of the more stable IPUs will increase with time, via the attrition of the less stable ones (survival of the fittest) and transfer of information among IPUs, both vertically (to progeny) and horizontally. This process engenders the key feature of Darwinian evolution, differential reproduction of genotypes, based on the feedback from the environment, transmitted through the phenotype.

   a. All 7 fundamental principles of the life-compatible universe (P1-P7) are involved in enabling evolution by natural selection. The very existence of units, on which selection can operate, hinges on the above-described phenomena, namely, self vs non-self discrimination of prebiotic IPUs (E1) and the emergence of shareable information storage (E6, E7). The crucial step to biology is the emergence of the link between the loss function, the existence of which is postulated by P1, on the one hand, and existence of the IPUs (P2, P3, P4, E1), on the other hand. Consumption of (limited) external resources (P5) entails competition between IPUs that share the same environment and turns mere shifts of the relative frequency into true "survival of the fittest". The ability of the IPUs to replicate (P6) and to expand their memory storage (genotype) (P7, E6, E7) provides them with access to hitherto unavailable degrees of freedom, making evolution an open-ended process rather than a quick, limited search for a local optimum.

   b. Evolution by natural selection is the central tenet of evolutionary biology and a key part of the NASA definition of life. An important note on definitions is due. We already referred to selection when discussing prebiotic evolution (E1); however, the term "natural (Darwinian) selection" is here reserved for the efficient form of selection that emerges with the replication of dedicated information storage devices (P6, E6). Differential reproduction, whereby the environment provides feedback on the success of genotypes, while acting on phenotypes, turns into Darwinian "survival of the fittest" in the presence of competition. When IPUs depend on environmental resources, such competition inevitably arises, except in the unrealistic case of unlimited supply (46). With the onset of Darwinian evolution, the system can be considered to cross the threshold from pre-life to life, so that all that follows belongs in the realm of biology (63, 64). The evolutionary process is naturally represented by movement of an evolving IPU in a genotype space, where proximity is defined by similarity between the distinct genotypes and transitions correspond to elementary evolutionary events, that is, mutations, in the most general sense (65). For any given environment, fitness - that is, a measure of the ability of a genotype to produce viable offspring - can be defined for each point in the genotype space, forming a multidimensional fitness landscape (55, 56). Selection creates a bias for preferential fixation of mutations that increase fitness, even if the mutations themselves occur in an entirely random manner. The existence of variation (mutations, genotypic changes, some of which have phenotypic effects) is traditionally listed as a component of the key triad (heredity, variability, selection), which comprise the conceptual foundation of evolutionary biology (47). Mutations are rare on the human timescale, so that realizing their importance was a major intellectual feat of the 19th century, primarily, attributable to Darwin (46). From the 21st century perspective, variability can be largely taken for granted because perfect information copying is theoretically impossible at non-zero temperatures and without access to unlimited resources (4).

E9. **Parasitism.** Parasites and host-parasite coevolution are ubiquitous across biological systems at multiple levels of organization and are both intrinsic to and indispensable for the evolution of life.

a. The flexibility of life-compatible systems (P5, P6) and the symmetry breaking in the information flow (P7), combined with the inherent tendency of life to diversify (E5), lead to a situation where parts of the system settle on a parasitic state, that is, scavenging information and matter from the host, without making a positive contribution to its fitness.
b. From the biological perspective, parasites evolve to minimize their direct interface with the environment, and conversely maximize their interaction with the host; in other words, the host replaces most of the environment for the parasite. Parasites inevitably emerge and persist in biological systems because of two reasons: i) the parasitic state is reachable via an entropy-increasing step, and therefore, is highly probable (15, 66) and ii) highly efficient anti-parasite immunity is costly (67). The cost of immunity reflects another universal trade-off analogous to the trade-off between information transfer fidelity and energy expenditure: in both cases, infinite amount of energy is required to reach a zero error rate or a parasite-free state. From a complementary standpoint, parasites inevitably evolve as cheaters in the game of life that scavenge resources from the host (or in other words, exploit the host as a resource) without expending energy on resource production. Short-term, parasites reduce the host fitness by both direct drain on its resources and by a plethora of indirect effects including the cost of defense. However, in a longer-term perspective, parasites comprise a reservoir for recruitment of new functions (especially, but far from exclusively, for defense) by the hosts (13, 14). The host-parasite relationship can evolve towards transition to a mutually beneficial, symbiotic lifestyle that can further progress to mutualism and, in some cases, complete integration as exemplified by the origin of essential endosymbiotic organelles in eukaryotes, mitochondria and chloroplasts (68-70). Parasites emerge at similar levels of biological organization (organisms parasitizing on other organisms) or across levels (genetic elements parasitizing on organismal genomes or cell clones parasitizing on multicellular organisms).

E10. **Programmed death.** Programmed (to a various degree) death is an intrinsic feature of life.

a. Replication and elimination of information processing units (P6) and utilization of additional degrees of freedom (P5) form the foundation for the phenomenon of programmed death. At some levels of organization (for example, intra-genomic), the ability to add and eliminate units (such as genes) for the benefit of the higher-level systems (such as organisms) provides an obvious path of optimization. The elimination of units could be, in principle, completely random, but selection (E8) generates a sufficiently strong feedback to facilitate and structure the loss process (for example, purging low-fitness genes via homologous recombination or altruistic suicide of infected or otherwise impaired cells). The same forces operate at all levels of organization and selection (P4). In particular, if population-level or kin-level selection is sufficiently strong, mechanisms for altruistic death of individual organisms can be fixed in evolution (71, 72).
b. Programmed death is a prominent case of minimization of the higher-level loss function (such as fitness of the whole organism) at the cost of increasing the lower-level loss function (such as survival of individual cells). Although (tightly controlled) programmed cell death was originally discovered in multicellular organisms and has been thought to be limited to these complex life forms, altruistic suicide now appears to be a universal biological phenomenon (72-74). Moreover, many ecological and evolutionary phenomena, such as the choice between the K- and r-strategies of selection, are strongly lifespan-dependent (75, 76). Therefore, evolutionary tuning of the genetically determined lifespan (a "weak" form of programmed death) could also play an important role in organismal evolution.

To conclude this section, which we titled "Fundamental evolutionary phenomena", deliberately omitting "biological", it seems important to note that phenomena E1-E7 are generic, applying to all learning systems, including purely physical and prebiotic ones. However, the onset of natural selection (E8) marks the origin of life, so that E8-E10 belong in the realm of biology.

## 4. Optimization and scale separation in evolving systems

In the previous sections, we formulated the 7 fundamental principles of evolution P1-P7 and then argued that most, if not all key evolutionary phenomena E1-E10 can be interpreted and analyzed in the context of these principles, and apparently, derived from the latter. The next step is to formulate a mathematical framework that would be consistent with the fundamental principles and thus would allow us to model evolutionary phenomena either analytically or numerically. For concreteness, the proposed framework is based on a mathematical model of artificial neural networks (77, 78), but we first start by formulating a general optimization approach in a form suitable for modeling biological evolution.

We are interested in a the broadest class of optimization problems, where the loss (or cost) function $H(\mathbf{x}, \mathbf{q})$, is minimized with respect to some trainable variables,

$$\mathbf{q} = \left(\mathbf{q}^{(c)}, \mathbf{q}^{(a)}, \mathbf{q}^{(n)}\right), \quad (4.1)$$

for a given training set of non-trainable variables,

$$\mathbf{x} = \left(\mathbf{x}^{(o)}, \mathbf{x}^{(e)}\right). \quad (4.2)$$

Near a local minimum, the first derivatives of the average loss function with respect to trainable variables $\mathbf{q}$ are small, and the depth of the minimum usually depends on the second derivatives. In particular, the second derivative can be either large, for the effectively constant degrees of freedom, $\mathbf{q}^{(c)}$, or small, for adaptable degrees of freedom, $\mathbf{q}^{(a)}$, or near zero, for symmetries or neutral directions $\mathbf{q}^{(n)}$. The separation of the neutral directions $\mathbf{q}^{(n)}$ into a special class of variables simply means that some of the trainable variables can be changed without affecting the learning outcome, that is, the value of the loss function. Put another way, neutral changes are always possible. The neutral directions $\mathbf{q}^{(n)}$ are the fastest-changing among the trainable variables because fluctuations resulting in their change are, in general, fully stochastic. On the other end of the spectrum of variables, even minor changes to the effectively constant variables $\mathbf{q}^{(c)}$ compromise the entire learning (evolution) process, that is, result in substantial increase of the loss function value; these variables correspond to deep minima of the loss function. When the basin of attraction of a minimum is deep and narrow, the system stays in its bottom for a long time, and then, to describe such a state, it is sufficient to use discrete information (that is, to indicate that the system stays in a given minimum) rather than to list all specific values of the coordinates in a multidimensional space.

In a generic optimization problem, the dynamics of both trainable and non-trainable variables involves a broad distribution of characteristic time scales $\tau$, and switching between scales is equivalent to switching between different frequencies or, in the context of biological evolution, between different levels of organization. For any fixed $\tau$, all variables can be partitioned into three classes depending on how fast they change with respect to the specified time scale:

- fast-changing non-trainable variables that characterize an organism ($x^{(o)}$) and its environment ($x^{(e)}$), and change on time scales $\ll \tau$.

- intermediate-changing adaptable variables $q^{(a)}$ or neutral directions $q^{(n)}$ that change on time scales $\sim \tau$

- slow-changing variables, which are the degrees of freedom $q^{(c)}$ that have already been well trained and are effectively constant (at or near equilibrium), only changing on time scales $\gg \tau$.

As it will become evident shortly, the separation of these three classes of variables and interactions between them are central to the evolution and selection on all levels of organization, resulting in pervasive multilevel learning and selection.

Depending on the considered time-scale $\tau$ (or as a result of environmental changes), the same dynamical degree of freedom can be assigned to different classes of variables, that is, $x^{(o)}, x^{(e)}, q^{(c)}, q^{(a)}$ or $q^{(n)}$. For example, on the shortest time scale, which corresponds to the lifetime of an individual organism (one generation), the adaptable variables are the phenotypic traits that quickly respond to environmental changes, whereas the slowest, near constant variables are the genomic sequences (genotype) that change minimally if at all. On longer time scales, corresponding to thousands or millions of generations, fast evolving portions of the genome become adaptable variables, whereas the conserved core of the genome remains in the near constant class (52). Analogously, the neutral directions correspond either to non-consequential phenotypic changes or to neutral genomic mutations, depending on the time scale. Indeed, in evolutionary biology, it is well known that the overwhelming majority of the mutations are either deleterious and are therefore eliminated by purifying selection, or (nearly) neutral and thus can be either lost or fixed via drift (79, 80). However, when the environment changes, or under the influence of other mutations, some of the neutral mutations can become beneficial (a genetic phenomenon known as epistasis, which is pervasive in evolution (81,82)), and in their entirety, neutral mutations form the essential reservoir of variation available for adaptive evolution (83). Furthermore, even which variables are classified as non-trainable ($x$), depends on the time scale $\tau$. For example, if a learning system was trained for a sufficiently long time, some of the trainable variables $q^{(a)}$ or $q^{(n)}$ might have already equilibrated, and become non-trainable.

## 5. The neural network framework

Now that we described an optimization problem that is suitable for modeling evolution of organisms (or populations of organisms), we can construct a mathematical framework to solve such optimization problems. For this purpose, we employ a mathematical theory of artificial neural networks (77, 78), which is simple enough to perform calculations, while being consistent with all of the fundamental principles (P1-P7), and thus can be used for modeling evolutionary phenomena (E1-E10). We first recall a general framework of the theory of neural networks.

Consider a learning system represented as a neural network, with the state vector described by trainable variables $q$ (which is a collective notation for weight matrix $\hat{w}$ and bias vector $b$) and non-trainable variables $x$ (which describe the current state vector of individual neurons). In the biological context, $x$ collectively represent the current state of the organism $x^{(o)}$ and of its environment $x^{(e)}$, and $q$ determine how $x$ changes with time, in particular, how the organism reacts to environmental challenges. The non-trainable variables are modeled as changing in discrete time-steps

$$x_i(t+1) = f_i\left(\sum_j w_{ij} x_j(t) + b_i\right) \quad (5.1)$$

where $f_i(y)$'s are some non-linear activation functions (for example, hyperbolic tangent or rectifier activation functions). The trainable variables are modeled as changing according to the gradient descent (or stochastic gradient descent) algorithm

$$q_i(t+1) = q_i(t) - \gamma \frac{\partial H(\mathbf{x}(t), \mathbf{q}(t))}{\partial q_i} \quad (5.2)$$

where $\gamma$ is the learning rate parameter and $H(\mathbf{x}, \mathbf{q})$ is a suitably defined loss function (see Eqs. (5.3) and (5.4) below). In other words, $\mathbf{q}$ are "gross", or "main" variables, which determine the rules of dynamics, and the dynamics of all other variables $\mathbf{x}$ is governed by these rules, per equation (5.1). In the biological context, equation (5.1) represents fast, often stochastic environmental changes, and the corresponding fast reaction of organisms at the level of the phenotype, whereas (5.2) reflects slower learning dynamics of evolutionary adaptation, via changes in the intermediate, adaptable variables, that is, the variable portion of the genome. The main learning objective is to adjust the trainable variables such that the average loss function is minimized subject to boundary conditions (also known as the training dataset), which in our case is modeled as a time sequence of the environmental variables.

For example, on a single generation time scale, the fast-changing variables represent the environment $\mathbf{x}^{(e)}$ and non-trainable variables associated with organisms $\mathbf{x}^{(o)}$, the intermediate-changing variables represent adaptive $\mathbf{q}^{(a)}$ and neutral $\mathbf{q}^{(n)}$ phenotype changes, and the slow-changing variables $\mathbf{q}^{(c)}$ represent the genotype (Figure 1).

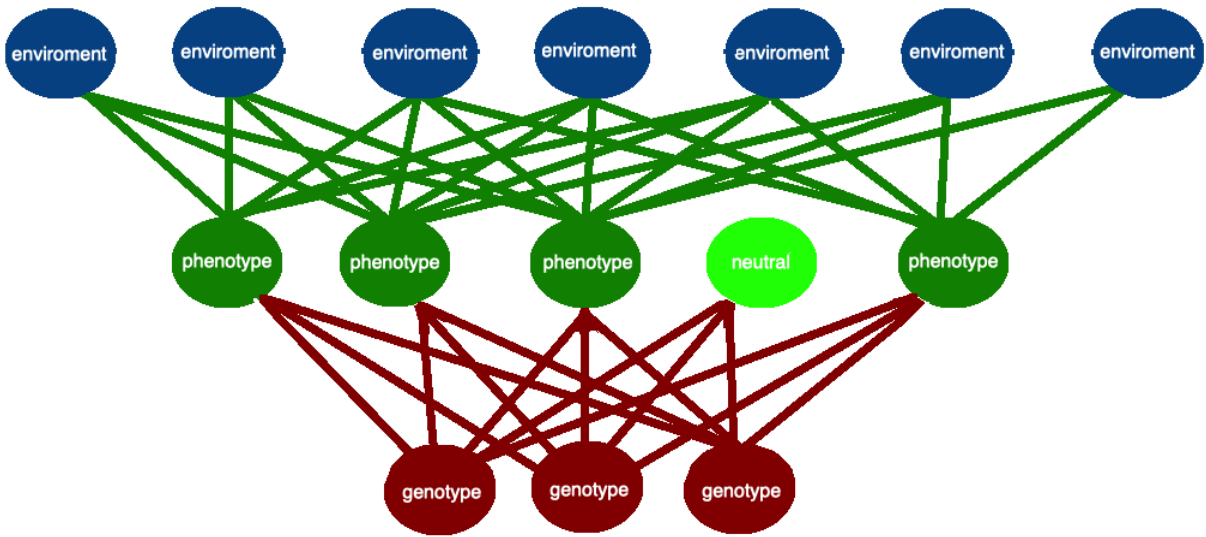

**Figure 1**. **Neural network with three layers.**

Non-trainable environmental variables (blue nodes), non-trainable organism variables (red and green nodes); trainable, intermediate-changing, adaptive phenotypic variables (dark green nodes and links); trainable, slow-changing (near constant), genotype variables (red nodes/links); and neutral variables (light green node).

The temporal scale separation in biology is readily apparent in all organisms. Indeed, consequential changes in the environment $\mathbf{x}^{(e)}$ often occur on the scale of milliseconds to seconds, triggering physical changes within organisms $\mathbf{x}^{(o)}$ at matching timescales. In response, individual organisms respond with phenotypic changes, both adaptive $\mathbf{q}^{(a)}$ and neutral $\mathbf{q}^{(n)}$ on the scale of minutes to hours, exploiting their genetically encoded phenotypic plasticity. A paradigmatic example is induction of bacterial operons in response to a change in the chemical composition of the environment, such as the switch from glucose to galactose as the primary nutrient (84, 85). In contrast, changes in the genome $\mathbf{q}^{(c)}$ take much longer. Mutations typically occur at rates of about 1 to 10 per genome replication cycle (86), which for unicellular organisms, is the same as a generation comprising from about an hour to hundreds or even thousands of hours. However, fixation of mutations, which represents an evolutionarily stable change at the genome level, typically takes many generations and thus always occurs orders of magnitude slower than phenotype changes. Accordingly, on this time scale, any changes in the genome represent the third layer in the network, the slowly changing variables.

To specify a microscopic loss function that would be appropriate for describing evolution, and thus, give a specific form to the fundamental principle P1, we first note that adaptation to the environment is more efficient, that is, the loss function value is smaller, for a learning system, such as an organism, that can predict the state of its environment with a smaller error. Then, the relevant quantity is the so-called "boundary" loss function defined as the sum of squared errors,

$$H_e(\mathbf{x}, \mathbf{q}) \equiv \frac{1}{2} \sum_{i \in \mathbb{E}} \left( x_i^{(e)} - f_i\left(\mathbf{x}^{(o)}, \mathbf{q}\right) \right)^2 \quad (5.3)$$

where the summation is taken only over the boundary (or environmental) non-trainable variables. It is helpful to think of the boundary loss function as the mismatch between the actual state of the environment and the state that would be predicted by the neural network if the environmental dynamics was switched off. In neuroscience, boundary loss is closely related to the surprise (or prediction error) associated with predictions of sensations, which depend on an internal model of the environment (87). In machine learning, boundary loss functions are most often used in the context of supervised learning (36), and in biological evolution, the "supervision" comes from the environment, which the evolving system, such as an organism or a population, is learning to predict.

Another possibility for a learning system is to search for the minimum of the "bulk" loss function, which is defined as the sum of squared errors over all neurons

$$H(\mathbf{x}, \mathbf{q}) = \frac{1}{2} \sum_i \left( x_i - f_i\left(\mathbf{x}^{(o)}, \mathbf{q}\right) \right)^2 . \quad (5.4)$$

The bulk loss function assumes extra cost incurred by changing the states of organismal neurons, $\mathbf{x}^{(o)}$, that is, rewarding stationary states. In the limit of a very large number of environmental neurons, the two loss functions are indistinguishable, $H(\mathbf{x}, \mathbf{q}) \approx H_e(\mathbf{x}, \mathbf{q})$, but bulk loss is easier to handle mathematically (the details of boundary and bulk loss functions are addressed in Ref. (36)).

More generally, in addition to the "kinetic" term (5.4), the loss function can include a "potential" term $V(\mathbf{x}, \mathbf{q})$

$$H(\mathbf{x}, \mathbf{q}) = \frac{1}{2} \sum_i \left(x_i - f_i\left(\mathbf{x}^{(o)}, \mathbf{q}\right)\right)^2 + V(\mathbf{x}, \mathbf{q}). \quad (5.5)$$

The kinetic term in (5.5) reflects the ability of organisms $\mathbf{x}^{(o)}$ to predict the changes in the state of the given environment $\mathbf{x}^{(e)}$ over time, whereas the potential term reflects their compatibility with a given environment and hence the capacity to choose among different environments.

In the context of biological evolution, Malthusian fitness $\varphi$ is defined as the expected reproductive success of a given genotype, that is, the rate of change of the prevalence of the given genotype in an evolving population (88). However, in the context of the theory of learning, the loss function must be identified with the additive fitness, that is,

$$H(\mathbf{x}, \mathbf{q}) = -T \log \varphi(\mathbf{x}, \mathbf{q}). \quad (5.6)$$

At the level of a microscopic description of learning, the proportionality constant is unimportant, but as we argue in detail in the accompanying paper (89), in the description of the evolutionary process from the point of view of thermodynamics, $T$ plays the role of "biological temperature".

Given a concrete mathematical model of neural networks, one might wonder if all fundamental principles of evolution (P1-P7) can be derived from this model. Such derivation would comprise additional evidence supporting the claim that the entire universe can be adequately described as a neural network (37). Clearly, the existence of a loss function P1 follows automatically because learning of any neural network is always described relative to a specified loss function (see Eqs. (5.4) or (5.5)). The other 6 principles also seem to naturally emerge from the learning dynamics of neural networks. In particular, the hierarchy of scales P2 and frequency gaps P3 are generic consequences of the learning dynamics, whereby a system that involves a wide range of variables changing at different rates is attracted towards a self-organized critical state of slow-changing trainable variables (90). Additional gaps between levels of organization are also expected to appear through phase transitions as becomes apparent in the thermodynamic description of evolution we develop in the accompanying paper (89). Renormalizability P4 is a direct consequence of the second law of learning (36), according to which entropy of a system (and consequently, complexity of neural network or rank of its weight matrix) decreases with learning. This phenomenon was observed in neural network simulations (36) and is the exact type of dynamics that can make the system renormalizable even if it started off as a highly entangled (large rank of weight matrix), non-renormalizable neural network. The extension P5 and replication P6 principles simply indicate that additional variables can lead to either increase or decrease in the value of the loss function (91). It is also important to note that in neural networks, an additional computational advantage ("quantum advantage") can be achieved if the number of information processing units can vary (92). Therefore, to achieve such an advantage, a system must learn how to replicate and eliminate its IPUs (P6). Finally, in Section 7, we illustrate how Fourier transform (or more generally, wavelet transform) of the environmental degrees of freedom can be used for learning the environment and how the inverse transform can be used for predicting it. Thus, to be able to predict the environment (and hence to be competitive), any evolving system must learn the mechanism behind such asymmetric information flow P7.

### 6. Multilevel learning

In the previous sections, we argued that the learning process naturally divides all the dynamical variables into three distinct classes, fast-changing, $\mathbf{x}^{(o)}$ and $\mathbf{x}^{(e)}$, intermediate-changing, $\mathbf{q}^{(a)}$ and $\mathbf{q}^{(n)}$ ($\mathbf{q}^{(n)}$ being faster than $\mathbf{q}^{(a)}$), and slow-changing ones, $\mathbf{q}^{(c)}$ (Fig. 1). Evidently, this separation of variables depends on the time-scale $\tau$, during which the system is observed, and variables migrate between classes when $\tau$ is increased or decreased (Fig. 2). The longer the time $\tau$, the more variables reach equilibrium and therefore can be modeled as non-trainable and fast-changing, $\mathbf{x}^{(e)}$, and the fewer variables remain slowly-varying and can be modeled as effectively constant $\mathbf{q}^{(c)}$. In other words, many variables that are nearly constant at short time scales migrate to the intermediate class at longer time-scales, whereas variables from the intermediate class migrate to the fast class.

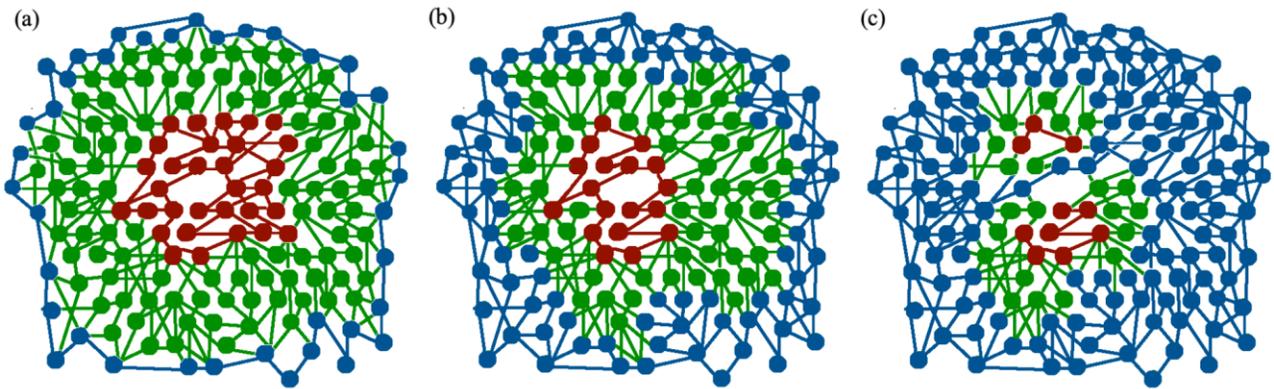

**Figure 2**. Separation of variables in a learning system depending on the time scale.

Three states of a learning system are shown, with fast-changing environmental variables (blue nodes and links), intermediate-changing trainable variables (green nodes and links) and slow-changing trainable variables (red nodes and links) observed on three different time-scales, from the shortest to longest: (a) $\rightarrow$ (b) $\rightarrow$ (c).

In biological terms, if we consider learning dynamics on the time-scale of a lifetime of an organism, then, $\underline{\mathbf{q}^{(c)}}$ and ($\underline{\mathbf{q}^{(a)}}$ or $\mathbf{q}^{(n)}$), represent the genotype and phenotype variables, respectively, but on much longer time-scales of multiple generations, the learning dynamics of populations (or communities) of organisms becomes relevant. On such time scales, the genotype variables acquire dynamics, with purifying and positive selection getting into play, whereas the phenotype variables progressively equilibrate. There is a clear connection between the learning dynamics, including that in biological systems, and renormalizability of physical theories (P4). Indeed, from the point of view of an efficient learning algorithm, the parameters controlling the learning dynamics, such as effective learning (or information processing) rate $\gamma$, can vary from one time-scale to another (for example, from individual organisms to populations or communities of organisms), but the general principles as well as specific dependencies captured in the equations above that govern the learning dynamics on different time-scales remain the same. We refer to this universality of the learning process on different time scales and partitioning of the variables into temporal classes as multilevel learning.

More precisely, multilevel learning is a property of learning systems, which allows for the basic equations of learning, such as (5.4), to remain the same on all levels of organization, but for the parameters, which describe the dynamics, such as $\gamma(\tau)$, to depend on the level, or on the time-scale $\tau$. For example, if the effective learning (or information processing) rate $\gamma(\tau)$ decreases with time-scale $\tau$, then the local processing time, which depends on $\gamma(\tau)$, runs differently for different trainable variables: slower for slow-changing variables (or larger $\tau$), and faster for fast-changing ones (or smaller $\tau$). For such a system, the concept of global time (that is, the same time for all variables) becomes irrelevant and should be replaced by the proper, or local time, which is defined for each scale $\tau$ separately,

$$t_\tau \propto \gamma(\tau)\, t. \quad (6.1)$$

This effect closely resembles the time dilation phenomena in physics, except that in special and general relativity, time dilation is linked with the possibility of movement between slow and fast clocks (or variables) (93). To illustrate the biological analog of time dilation and to understand the role it plays in biology, consider only two types of variables: slow-changing and fast-changing. Then, the slow variables should be able to "outsource" certain computational tasks to faster variables. Because the local clock for the fast-changing variables runs faster, the slow-changing variables can take advantage of the fast-changing ones to accelerate computation, which would be rewarded by evolution. The flow of information between slow-changing and fast-changing variables in the opposite direction is also beneficial because the fast-changing variables can use the slow-changing variables to store useful information for future retrieval, that is, the slow variables function as long-term memory. In the next section, we show that such cooperation between slow- and fast-changing variables, which is a concrete manifestation of the principle P7, corresponds to a crucial biological phenomenon known as the Central Dogma of molecular biology (59).

## 7. Generalized Central Dogma of Molecular Biology

In terms of learning theory, the two directions of the asymmetric information flow (P7) represent, respectively, learning the state of the environment and predicting the state of the environment, based on the results of learning. For learning, information is passed from faster variables to slower variables, and for predicting, information flows in the opposite direction, from slower variables to the faster ones. A more formal analysis of the asymmetric information flows (or what could be called the generalized Central Dogma) can be carried out by forward-propagation (from slow variables to fast variables) and back-propagation (from fast variables to slow variables) of information within the framework of the mathematical model of neural networks developed in the previous sections (Fig. 3).

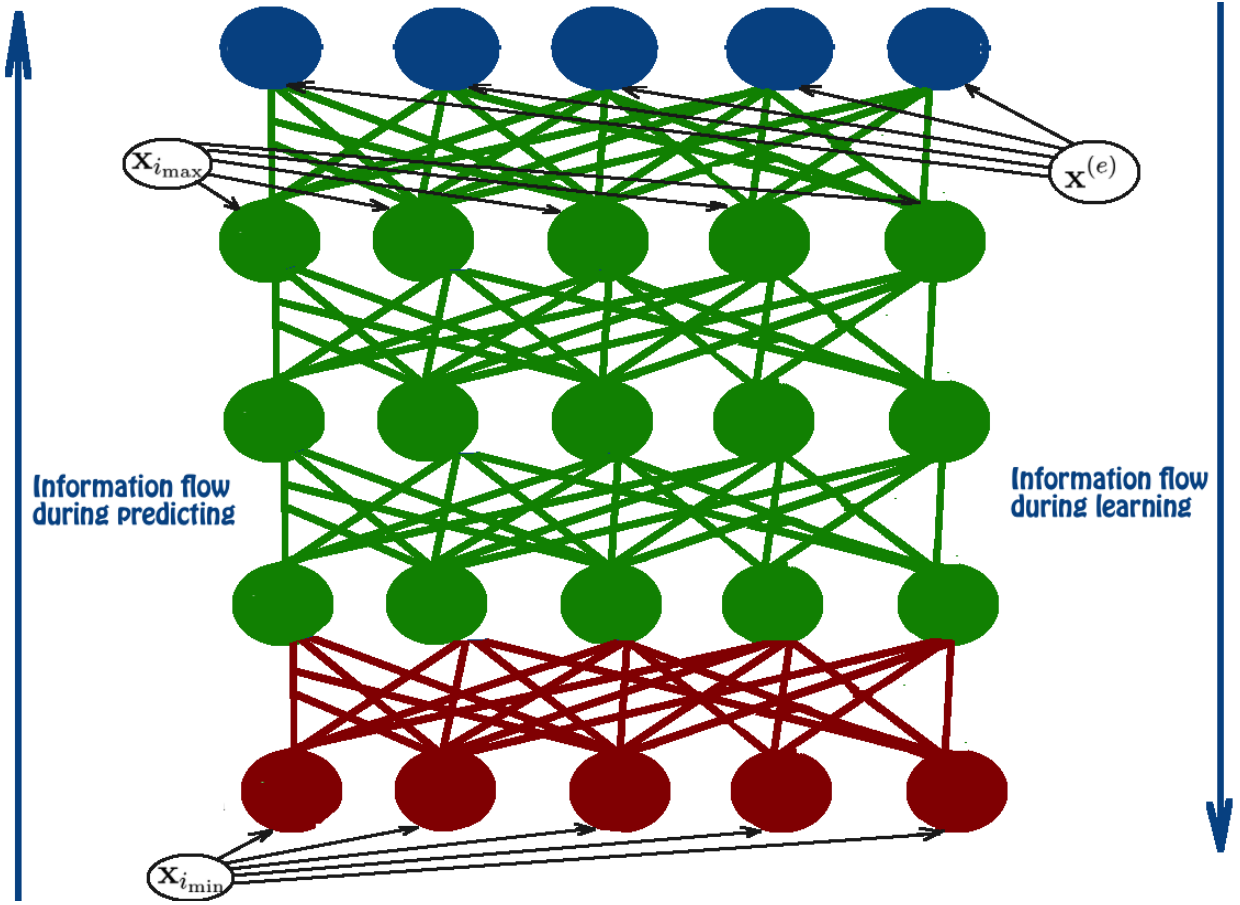

**Figure 3**. Asymmetrical information flow involved in learning and predicting the environment by evolving systems.

Consider non-trainable environmental variables that change continuously with time $\mathbf{x}^{(e)}(t)$, while the learning objective of an organism is to predict $\mathbf{x}^{(e)}(t)$ at time $t > \tau$, given that it was observed for time $0 < t < \tau$. Thus, the organism has to extrapolate the function $\mathbf{x}^{(e)}(t)$ for $t > \tau$ and to do so, it should be able to store and retrieve the values of the Fourier coefficients

$$\mathbf{q}^i = \frac{1}{\tau} \int_0^\tau dt \, \mathbf{x}^{(e)}(t) e^{-i 2\pi f_i t}. \quad (7.1)$$

or, more generally, wavelet coefficients

$$\mathbf{q}^i = \frac{1}{\tau} \int_0^\tau dt \, W_i(\tau - t) \, \mathbf{x}^{(e)}(t) e^{-i 2\pi f_i t} \quad (7.2)$$

for suitably defined window functions $W_i$. Then, a prediction could be made by extrapolating $\mathbf{x}^{(e)}(t)$ using the inverse transformation

$$\mathbf{x}^{(e)}(\tau + \delta) \approx 2 \sum_{i=i_{min}}^{i_{max}} \text{Re}\left(\mathbf{q}^i e^{i2\pi f_i \delta}\right) \quad (7.3)$$

for some $\delta > 0$, which is not too large compared to $\tau$. However, in general, the total number of (Fourier or wavelet) coefficients $\mathbf{q}^i$'s would be countably infinite. Therefore, any finite size organism has to "decide", which frequencies to observe (and remember) and which ones to filter out (and "forget").

Let us assume that the organism "decided" to only observe/remember discrete frequencies

$$f_{\min} \equiv f_{i_{\min}}, ..., f_{i_{\max}} \equiv f_{\max} \quad (7.4)$$

and forget everything else. Then, to predict the state of the environment (7.3) and, as a result, minimize the loss function (5.5), the organism should be able to store, retrieve and adjust information about coefficients $\mathbf{q}^i$'s in some adaptable trainable variables $\mathbf{q}^{(a)}$.

Given this simple model, we can study the flow of information between different non-trainable variables of the organism $\mathbf{x}^{(o)}$. To this end, it is convenient to organize the variables as

$$\mathbf{x} = \left(\mathbf{x}^{(o)}, \mathbf{x}^{(e)}\right) = \left(\mathbf{x}_{i_{\min}}, ..., \mathbf{x}_{i_{\max}}, \mathbf{x}^{(e)}\right) \quad (7.5)$$

where

$$\mathbf{x}_i(\tau + \delta) \approx 2 \sum_{k=i_{\min}}^{i} \text{Re}\left(\mathbf{q}^k e^{i2\pi f_k \delta}\right) \quad (7.6)$$

and assume that the relevant information about $\mathbf{q}^i$'s is stored in the adaptable trainable variables

$$\mathbf{q}^{(a)} = \left(\mathbf{q}^{i_{\min}}, ..., \mathbf{q}^{i_{\max}}\right). \quad (7.7)$$

In the estimate of $\mathbf{x}_i(\tau + \delta)$ in Eq. (7.6), all the higher frequency modes are assumed to average to zero as is often the case if we are only interested in the time scale $f_i^{-1}$. A somewhat better estimate can be obtained using, once again, the ideas of the renormalization group flow, following the fundamental principle P4. To make learning (and thus survival) efficient, truncation of the set of variables relevant for learning is crucial. The main point is that the higher frequency modes can still contribute statistically, and then, an improved estimate of $\mathbf{x}_i(\tau + \delta)$ would be obtained by appropriately modifying the values of the coefficients $\mathbf{q}^i$'s. Either way, in order to make an actual prediction, the organism should first calculate $\mathbf{x}_i(\tau + \delta)$ for small $f_i$ and then pass the result to the next level to calculate $\mathbf{x}_{i+1}(\tau + \delta)$ for larger $f_{i+1}$, and so on. Such computations can be described by a simple mapping

$$\mathbf{x}_{i+1}(\tau + \delta) = \mathbf{x}_i(\tau + \delta) + 2\,\text{Re}\left(\mathbf{q}^{i+1} e^{i2\pi f_{i+1} \delta}\right) \quad (7.8)$$

which can be interpreted as passage of data from one layer to another in a deep, multilayer neural network (Fig. 2). Eq. (7.8) implies that, during the predicting phase, relevant information only flows from variables encoding low frequencies to variables encoding high frequencies, but not in the reverse direction. In other words, in the process of predicting the environment, information propagates from slower variables to faster variables, that is, from genotype to phenotype, or from nucleic acids to proteins, hence the Central Dogma. Because only the fast variables change in this process, the prediction of the state of the environment is rapid, as it is indeed required to be for the organism survival. Conversely, in the process of learning the environment, information is back-propagated in the opposite direction, that is, from faster to slower variables. However, this back-propagation is not a microscopic reversal of the forward-propagation, but a distinct, much slower process (given that changes in slow variables are required) that involves mutation and selection.

Thus, the meaning of the generalized Central Dogma from the point of view of the learning theory – and our theory of evolution - is that slow dynamics (that is, evolution on a long time-scale) should be mostly independent of the fast variables. In less formal terms, slow variables determine the rules of the game, and changing these rules depending on the results of some particular games would be detrimental for the organism. Optimization within the space of opportunities constrained by temporally stable rules is advantageous compared to optimization without such constraints. The trade-off between global and local optimization is a general, intrinsic property of frustrated systems (E2). For the system to function efficiently, the impact of local optimization on the global optimization should be restricted. The separation of the long-term and short-term forms of memory through different elemental bases (nucleic acids vs proteins) serves this objective.

## 8. Discussion

In this work, we outline a theory of evolution on the basis of the theory of learning. The parallel between learning and the process of biological evolution is becoming obvious as soon as the mapping between the loss function and the fitness function is identified (Eq. (5.6)). Indeed, both processes represent movement of an evolving (learning) system on a fitness (loss function) landscape, where adaptive (learning), upward moves are most consequential although neutral moves are most common, and downward moves also occur occasionally. However, we go beyond the obvious analogy and trace a detailed correspondence between the essential features of the evolutionary and learning processes. Arguably, the most important, fundamental commonality between evolution and learning is the stratification of the trainable variables (degrees of freedom) into multiple classes that differ by the rate of change. At least, in complex environments, all learning is multilevel learning, and all selection relevant for the evolutionary process is intrinsically multilevel. This is a substantial deviation from the current mainstream narrative of evolutionary biology, in which multilevel selection remains a controversial subject, and in any case, is not generally considered to be the central evolutionary trend. However, the framework of evolution as learning developed here implies that evolution of biological complexity would be impossible without multilevel selection permeating the entire history of life. Under this perspective, the emergence of new levels of organization, in learning and in evolution, and in particular, MTE represent genuine phase transitions as previously suggested (42). Such transitions can be analyzed consistently only in the thermodynamic limit, which is addressed in detail in the accompanying paper (89).

The detailed correspondence between the key features of the processes of learning and evolution implies that this is not a simple analogy but rather a reflection of the deep unity of evolutionary processes occurring in the universe. Indeed, the separation of the relevant degrees of freedom into multiple temporal classes is ubiquitous in the universe, from composite subatomic particles, such as protons, to atoms, molecules, life forms, planetary systems and galaxy clusters. If we take the seemingly radical but actually straightforward and consistent view that the entire universe is a neural network (37), then, all these systems would be considered emerging from the learning dynamics. Furthermore, scale separation and renormalizability appear to be essential conditions for a universe to be observable. According to the evolution theory outlined here, any observable universe consists of entities that undergo learning, or synonymously, adaptive evolution, and actually, the universe itself is such an entity, in development of the concept of the world as a neural network (37). The famous dictum of Dobzhansky (94), thus, can and arguably should be rephrased as "Nothing in the world is comprehensible except in the light of learning."

Within the framework of this theory of evolution, the difference between life and non-living matter is one in degree of optimization rather than in kind. Crucially, any complex optimization problem can be addressed only with a stochastic learning algorithm, hence the ubiquity of selection. Origin of life can then be conceptualized within the framework of multilevel learning as we explicitly show in the accompanying paper (89). The point when life begins can be naturally associated with the emergence of a distinct class of slowly changing variables that are digitized and thus can be accurately replicated; these digital variables store and supply information for forward-propagation to predict the state of the environment. In biological terms, this focal point corresponds to the advent of replicators (genomes) that carry information on the operation of reproducers, within which they resided (95). This is also the point when natural (Darwinian) selection takes off (63). Our theory of evolution implies that this pivotal stage was preceded by evolution of "pre-life", which comprised reproducers that lacked genomes but nevertheless were learning systems that were subject to selection for persistence. The first replicators (RNA molecules) would evolve within these reproducers, perhaps, initially, as molecular parasites (E9), but subsequently, under selection for the ability to store, express and share information essential for the entire system. This crucial step greatly increased the efficiency of evolution/learning and provided for long term memory that persisted throughout the history of life, providing for the onset of natural selection and the unprecedented diversification of life forms. For learning to be efficient, the capacity of the system to add new adaptable variables is essential. In biological terms, this implies expandability of the genome, that is, ability to add new genes, which necessitated the transition from RNA to DNA as the genome substrate, given the apparent inherent size constraints on replicating RNA molecules. Another essential condition for efficient learning is information sharing which in the biological context corresponds to horizontal gene transfer. The essentiality of horizontal gene transfer at the earliest stages of life evolution is perceived as the cause of the universality of the translation machinery and genetic code in all known life forms (96).

The scenario of the origin of life within the encompassing framework of the presented theory of evolution, even if formulated in most general terms, implies that emergence of complexity commensurate with life is a general trend in the evolution of complex systems. At face value, this conclusion might seem to be at odds with the magnitude of complexification involved in the origin of life (suffice it to consider the complexity of the translation system (6)) and the uniqueness of this even, in the least, in the history of earth, and probably, on a much greater cosmic scale. Arguably, however, the origin of life can be conceptualized as an expected outcome of learning subject to the relevant constraints, such as the presence of the required chemicals in sufficient concentrations and more. Such constraints would make life a rare phenomenon but likely far from unique, on the scale of the universe.

Evidently, the analysis presented here and in the accompanying paper (89) is only an outline of a theory of evolution as learning. The details and implications including directly testable ones remain to be worked out.

**Author contributions**

V.V., Y.I.W., M.I.K. and E.V.K. conceptualized the project; V.V. developed the mathematical formalism; V.V. and E.V.K. wrote the manuscript with contributions from all authors.

**Acknowledgements**

E.V.K. is grateful to Dr. Purificacion Lopez-Garcia for essential discussions and critical reading of the manuscript. V.V. was supported in part by the Foundational Questions Institute (FQXi) and the Oak Ridge Institute for Science and Education (ORISE). Y.I.W. and E.V.K. are supported by the Intramural Research Program of the National Institutes of Health of the USA.